%% file: main.tex
\documentclass{article}
\usepackage{arxiv}
\usepackage[utf8]{inputenc}
\usepackage[pdftex]{graphicx}
\usepackage{mathtools}
\usepackage{amsmath}
\usepackage{array}
\usepackage{mathrsfs}
\usepackage{amssymb}
\usepackage{booktabs}
\usepackage{multirow}
\usepackage{tabularx}
\usepackage{tabulary}
\usepackage{dcolumn}
\usepackage{threeparttable}
\usepackage{bbm}
\usepackage{color, soul}
\usepackage{blindtext}
\usepackage{setspace}
\usepackage{geometry}
\usepackage{wrapfig}
\usepackage{lscape}
\usepackage{rotating}
\usepackage{epstopdf}
\usepackage[section]{placeins}
\usepackage{wasysym}
\usepackage[]{algorithm}
\usepackage{algpseudocode}
\usepackage{amsfonts}
\usepackage[toc,page]{appendix}
\usepackage{siunitx}
\usepackage{gensymb}
\usepackage{changes}
\usepackage{etoolbox}
\usepackage{multirow}
\usepackage{afterpage}
\usepackage{comment}
\usepackage{pgf}
\usepackage{siunitx}
\usepackage{standalone}
\usepackage{hyperref}
\usepackage[english]{babel}
\usepackage{fancyvrb}
\usepackage{esvect}
\usepackage{braket}
\usepackage{subcaption}

\DeclareMathOperator*{\argmin}{arg\,min}

\hypersetup{
	colorlinks,
	linkcolor={blue!75!black},
	citecolor={blue!75!black},
	urlcolor={blue!75!black}
}
\providecommand{\keywords}[1]
{
	\small	
	\textbf{\textit{Keywords:}} #1
}

\allowdisplaybreaks
\usepackage{tikz}
\usetikzlibrary{shapes,arrows,fit,positioning,calc}
\usepackage{qcircuit}
\usepackage[numbers,sort&compress]{natbib}
\tikzset{main node/.style={circle,draw=black,font=\sffamily\bfseries},edge_style/.style={draw=black,densely dashed},
}
\tikzset{
	>=stealth',
	punkt/.style={
		rectangle,
		rounded corners,
		draw=black, very thick,
		text width=8.5em,
		minimum height=2em,
		text centered},
	pil/.style={
		->,
		thick,
		shorten <=2pt,
		shorten >=2pt,}
}

\usepackage{metalogo}
\usepackage{dtklogos}
\usepackage{lipsum} 
\usepackage{fancyhdr}
\allowdisplaybreaks 

\title{Quantum Circuit Design Search}

\author{
	Mohammad Pirhooshyaran \\
	ISE, Lehigh University\\
	\texttt{mop216@lehigh.edu} \\
	\And 
	Tam\'{a}s Terlaky \\
	ISE, Lehigh University\\
	\texttt{terlaky@lehigh.edu}	\\
}

\begin{document}

	\maketitle

\begin{abstract}
This article explores search strategies for the design of parameterized quantum circuits. We propose several optimization approaches including random search plus survival of the fittest, reinforcement learning both with classical and hybrid quantum classical controllers and Bayesian optimization as decision makers to design a quantum circuit in an automated way for a specific task such as multi-labeled classification over a dataset. We introduce nontrivial circuit architectures that are arduous to be hand-designed and efficient in terms of trainability. In addition, we introduce reuploading of initial data into quantum circuits as an option to find more general designs. We numerically show that some of the suggested architectures for the Iris dataset accomplish better results compared to the established parameterized quantum circuit designs in the literature. In addition, we investigate the trainability of these structures on the unseen dataset Glass. We report meaningful advantages over the benchmarks for the classification of the Glass dataset which supports the fact that the suggested designs are inherently more trainable.
\end{abstract}


\section{Introduction}
In the era of noisy intermediate scale quantum (NISQ) \cite{preskill2018quantum} devices, variational quantum algorithms (VQA) \cite{mcclean2016theory} are promising tools to integrate classical machine learning methods with quantum frameworks. Parameterized quantum circuits (PQC) \cite{mitarai2018quantum} are introduced to tune parameters of a quantum circuit by leveraging classical optimization methods. Variational eigenvalue solvers \cite{peruzzo2014variational,kandala2017hardware} and quantum neural networks (QNN) \cite{farhi2018classification,grant2018hierarchical,verdon2018universal,skolik2020layerwise,beer2020training,schuld2020circuit} are among the many applications of PQC.

Analytical approaches towards calculating partial derivatives of the PQC outcome with respect to its parameters such as parameter-shift rule \cite{mitarai2018quantum, schuld2019evaluating, crooks2019gradients} pave the way for using gradient-based optimization methods to tune the circuit parameters.

However, recent studies identify barren plateaus and vanishing gradient problems related to the PQC training landscapes and cost-functions in QNNs \cite{mcclean2018barren, cerezo2020cost}. For a wide class of PQCs, random designs and/or defining cost functions in terms of global observables result in an exponentially vanishing partial gradients of the cost function with respect to the circuit parameters. Researchers suggest initialization \cite{grant2019initialization}, data reuploading \cite{perez2020data} strategies and introduce cost functions in terms of local observables \cite{cerezo2020cost} to alleviate the negative effect of vanishing gradient. However, finding more efficient circuit designs in terms of trainability and for parameterized quantum circuits has not been fully explored yet although early quantum circuit design studies date back to the late 90s \cite{williams1998automated} and early 2000s \cite{yabuki2000genetic}.

In many cases, some established designs are used for PQC framework consisting of layers with a fixed pair of parameterized rotation operation such as $R_y$ accompanied by a non-parameterized gate such as a controlled X or Z gates. \citet{ostaszewski2019quantum} introduce \texttt{Rotoselect} algorithm which sequentially optimizes the rotation operators and values in the PQC. Ansatz architecture search (AAS) is proposed by \citet{li2020quantum} via a novel Gibbs objective function. \citet{rattew2019domain} presents evolutionary algorithm to dynamically generate and optimize the expectation of an ansatz given a Hamiltonian. In their method, \texttt{MoG-VQE}, \citet{chivilikhin2020mog} utilize the non-dominated sorting genetic algorithm (NSGA-II) to optimize the topology of the variational ansatz itself. \citet{alexeev2020reinforcement} present a reinforcement learning (RL) method responsible for optimizing the quantum approximate optimization algorithm (QAOA) circuit parameters. Recently, \citet{zhang2020differentiable} brings classical neural architecture search (NAS) framework into quantum circuit design search. We, independently of this work, investigate the same paradigm. In this study, however, we focus on suggesting powerful designs in terms of trainability when the searching space is discrete and further extend the benchmark PQC modeling (see equation (\ref{eq:circuit})) by introducing data reuploading as an option.

We use three separate design search strategies. We implement a simple yet effective random search over the vast space of feasible designs combined with the survival of the fittest. Second, a reinforcement learning approach for quantum circuit design search (R-QCDS) is investigated. A policy gradient based controller is introduced due to the successful predecessor work in classical machine learning (ML) area \cite{zoph2016neural, pham2018efficient}. Moreover, for the first time, we investigate hybrid quantum classical policy gradient based controllers. Third, a black-box Bayesian optimization (BO) is presented. We assume discrete design search space in contrast to \citet{zhang2020differentiable} where use continuous differentiable search space and probabilistic models.

In this study, we aim to keep the number of parameters of PQCs fixed while trying to improve their test accuracy (loss) for the task of supervised ML. We discover and present many nontrivial circuit designs that are efficient enough to compete with a typical (rotation-controlled) designs in the literature. In addition, we tackle the task of multi-label classification by considering small to moderate datasets such as Iris and Glass. Glass dataset in particular consists of 6 separate labels. We do not limit the QCDS only to device-efficient operations that have already been applied successfully on real quantum devices. In other words, we are not necessarily concerned about the extra number of controlled gates or non-linear connections between qubits in a design but instead we suggest designs that are efficient in terms of their ability to be trained. We hope this shed some light on designing new circuits as well. The rest of the paper is as following. We discuss the problem and its background in the Section \ref{Theory}. Then, we present the search strategies in Section \ref{desgin}. The results will be reported in Section \ref{results} and at last we present some extra discussion and conclusion in Section \ref{Conclusion}. In addition, parts of the implementation plus successful designs are available at the paper's \href{https://github.com/mamadpierre/QCDS}{github repository}.


\section{Theory and Background}
\label{Theory}
In general, PQCs are favorable approaches because \cite{cerezo2020cost,farhi2018classification}: First, one can theoretically design a task-specific PQC; Second, a PQC can benefit from new advances in classical paradigms at least in two ways, either by outsourcing the parameter update routine to a classical optimization algorithm or by using a PQC as only a component of a larger design, involving other classical parts; Third, these methods are more resistant against quantum noise, because the circuit parameters can learn the effect of the noise to some degree during the training; Forth, the PQC volume may be independent of or logarithmic in the size of the dataset they are dealing with \cite{perez2020data}. That is, in contrast with many early quantum algorithms, such as Grover's algorithm which needs a depth of around $\sqrt{N}$, where $N$ is the size of the dataset, in PQC, the depth of the parameterized circuit (i.e., the number of layers of the quantum operators stacked upon each-other) and the width of the circuit (the summation of all the input and ancilla qubits) can be a constant or relatively small number compared to the size of the dataset.   

In this study, we consider the classification task of 

	\begin{equation} \min_{\theta} J\left(\theta\right) := E\left(\mathcal{L}\left(f\left(x;\theta \right),y\right)\right) \label{eq:model} \end{equation}
where the PQC function $f\left(x;\theta \right)$ is given as

	\begin{equation}  f\left(x;\theta \right) =  \braket{0|U\left(x;\theta\right)^{\dagger}H U\left(x;\theta\right)|0}.  \label{eq:circuit} \end{equation}
The quantum circuit $U\left(x;\theta\right)$ is parameterized over $\theta$ and receives $x$ as an input feature. $\mathcal{L}$ is an appropriate loss function which compares the output of the PQC function $f$ and the true labels $y$. The outer expectation is due to the stochasticity of the setting. This might come from the fact that we empirically evaluate $f\left(x;\theta \right)$ at random subsamples $x$ and never have access to the unknown distribution $X$ where the input $x$ is drown from or arise from some inherent noise. In addition, we have an implicit expectation over observable $H$ at the end of the PQC, $ \braket{\psi,H\psi}$ where $\ket{\psi} = \ket{U\left(x;\theta\right)|0}$. Unitary $U\left(x;\theta\right)$ can be written as a chain of unitary transformations \cite{mitarai2018quantum, cerezo2020cost,grant2019initialization}. We propose  
  
	\begin{equation} 
		U\left(x;\theta\right) = \prod_{l=0}^{L-1} \bigotimes_{i = 0}^{N-1} \left(T_i^l \left(x_i^0 \right) U_i^l\left(\theta_i^l \right) W_i^l \right)\label{eq:PQC} 
	\end{equation}
	
as a general design of the PQC, where $N$ is the number of qubits and $L$ is the number of the layers. Superscripts represent layers while subscripts represent qubits. Further, $x_i^0$ is a subset of input features $x$ where at the beginning has been injected for the first time into qubit $i$ of the circuit. $T_i^l \left(x_i^0 \right)$ is an option of reuploading $x_i^0$ for qubit $i$ at layer $l$. We assume, either $T_i^l \left(x_i^0 \right) = R_y\left(x_i^0 \right)$ meaning that for layer $l$ we reupload the initial features via a rotation gate around the $y$-axis, or $T_i^l \left(x_i^0 \right) = I$, where $I$ is an appropriate identity matrix meaning that we reject the feature reuploading for layer $l$.

We assume $U_i^l\left(\theta_i^l \right) = exp\left(-i\theta_i^l V_i^l\right)$, where $V_i^l$ is a Hermitian operator, particularly we consider Pauli matrices $\sigma_x$, $\sigma_y$ and $\sigma_z$ as possible choices for $V_i^l$. $ W_i^l$ are the nonparametric gates and we consider Pauli matrices, Hadamard, CNOT, CSWAP and Toffoli choices. Please refer to Figure \ref{IrisToGlassExamples} to see two examples of the above expressions.

Similar designs to the ones discussed in the literature \cite{cerezo2020cost,grant2019initialization} can be extracted as special cases of equation (\ref{eq:PQC}) by turning off the reuploading option, choosing $U_i^l\left(\theta_i^l \right) = R_y\left(\theta_i^0 \right)$ and a CZ as $ W_i^l$ for all qubits $i \in \{0,1, \dots N-1\}$ and $l \in \{0,1, \dots L-1\}$.     

We consider

	\begin{equation} 
		  H =  \bigotimes_{i \in \left[\bar{N}\right]} \sigma_z \bigotimes_{i \in \left[N\right] \setminus \left[\bar{N}\right]} I \label{eq:obseravble} 
	  \end{equation}
as the observable in equation (\ref{eq:circuit}), where $\left[N\right] = \{0,1,\dots,N-1\}$ and $\left[\bar{N}\right] = \{0,1,\dots,\bar{N}-1\}$ is the set of all qubits that will be measured. Note that $N$ and $\bar{N}$ (the number of input qubits and the number of output measurements) are not necessarily the same. The introduced observable lacks physical intuition 
or operational meaning like trace distance cost functions \cite{huang2019near,jones2018quantum} which inherently provide us with a bound on the expectation of some states difference or fidelity. But trace distance cost functions are very difficult to train due to the vanishing gradient phenomena \cite{cerezo2020cost} while the above observable provides reliable trainability results. In addition, the observable is suitable for classification purposes. The output of the PQC can directly go through a classical softmax to produce a probability distribution for the classification.

As long as Hermitian operators generating unitaries of the parametric gates, have two unique eigenvalues, which is the case for Pauli matrices, parameter-shift rule applies for partial derivative calculations \cite{schuld2019evaluating}. For the sake of presenting the partial derivatives of $f\left(x;\theta \right)$ with respect to its parameters, we modify equation (\ref{eq:PQC}) as 

	\begin{equation} 
	U\left(x;\theta\right) = \prod_{l=0}^{L-1} \prod_{i = 0}^{N-1} \left(\tilde{T}_i^l \left(x_i^0 \right) \tilde{U}_i^l\left(\theta_i^l \right) \tilde{W}_i^l \right)\label{eq:PQCModified} 
\end{equation}

where 
\begin{align}
\tilde{T}_i^l 	& =  \bigotimes_{j \in \left[N\right] \setminus i} I \bigotimes T_i^l \nonumber\\
\tilde{U}_i^l	& =  \bigotimes_{j \in \left[N\right] \setminus i} I \bigotimes U_i^l \nonumber\\
\tilde{W}_i^l	& =  \bigotimes_{j \in \left[N\right] \setminus i} I \bigotimes W_i^l.  \label{eq:RevisedOperators}
\end{align}
 
Then, we have  
	\begin{equation} 
		 \dfrac{\partial f\left(x;\theta \right)}{\partial \theta_j^k} =  i\braket{0|\tilde{U}_{-}^{\dagger}\left[V_j^k, \tilde{U}_{+}^{\dagger} H \tilde{U}_{+}\right]\tilde{U}_{-}|0}  \label{eq:partial}
	  \end{equation}

where $\tilde{U}_{-} = \prod_{l=0}^{k} \prod_{i = 0}^{j-1} \left(\tilde{T}_i^l \left(x_i^0 \right) \tilde{U}_i^l\left(\theta_i^l \right) \tilde{W}_i^l \right)$ and $\tilde{U}_{+} = \prod_{l=k}^{L-1} \prod_{i = j}^{N-1} \left(\tilde{T}_i^l \left(x_i^0 \right) \tilde{U}_i^l\left(\theta_i^l \right) \tilde{W}_i^l \right)$.

The PQC architecture can be used as a part of a larger architecture including other classical and/or quantum parts. The input feature $x$ can be the output of another training architecture. Particularly, many large machine learning datasets can be down-scaled to a feature size that today's quantum circuits can handle. Moreover, instead of going through the loss function, the output of PQC $f\left(x;\theta \right)$ can continue to be fed into another architecture as an input feature. As long as we consider PQC as the last architecture, we assume the number of qubits involved in the output measurements ($\bar{N}$) equals to the number of classes and then the output of the PQC goes through the negative log-likelihood loss function. In the following section we discuss the approaches we use to search for the PQC architectures.


\section{Design Search}
\label{desgin}

The QCDS problem can be modeled as an optimization problem
\begin{equation}
	\displaystyle	\argmin_{d} \ \displaystyle \min_{\theta} J\left(\theta\right); \\\ \textrm{s.t.} \  d\in D,  
\end{equation}

where $J\left(\theta\right)$ was introduced in equation (\ref{eq:model}) and $d$ is a design drawn from $D$, the set of all the feasible designs, to identify the exact operations taken by $U\left(x;\theta\right)$ in equation (\ref{eq:PQC}). As the closest work to this study, \citet{zhang2020differentiable} introduce the phrase, ``quantum architecture search (QAS).'' We, however, suggest keep using the phrase ``quantum circuit design search (QCDS)'' because similar phrases are already established in the early circuit design publications \cite{williams1998automated, yabuki2000genetic, dutta2018quantum, raeisi2012quantum}.

QCDS consists of three parts: establishing the search space $D$, proposing search strategy, and performance metric. In this study, we assume single qubit search space. In other words, we look at PQC designs from a micro-level point of view. Each qubit can accept or reject the reuploading of the data it observes at the beginning of the circuit. It can then choose any rotation in $\{R_x, R_y, R_z\}$ and at last it can choose a nonparametric gate in \{H, $\sigma_x, \sigma_y, \sigma_z$, CNOT, CSwap, Toffoli, CZ\}. For a simple layer of 4 qubits the search space passes more than 5 million designs and by stacking-up layers of qubits, we make $|D|$ exponentially larger. Another approach for building a QCDS search space is to consider blocks of operations governing on more than one qubit. But this approach requires strong and efficient quantum blocks which brings us back to the micro-level approach. 

For the search strategy we assume three methods, reinforcement learning, random search and Bayesian optimization. We go through them in the rest of this section. For the performance metric we consider either the accuracy of the validation dataset or its loss value.
    
\subsection{Reinforcement learning for QCDS}
\label{RQCDS}
In this section we focus on reinforcement learning for quantum circuit design search (R-QCDS). We introduce an environment similar to the classical machine learning work \cite{zoph2016neural, pham2018efficient} in which a controller and a PQC interact with each other to find the best design for the PQC. We propose a deep neural network (DNN) as the controller. At each time-step $t$ controller outputs a policy (joint distribution) $\Pi_t(w)$ over all the decisions to be made, parameterized by $w$ the weights of the DNN. Then, a sample is drwan from $\Pi_t(w)$ as the suggested design. The PQC trains based on the design over training set and then reports back the accuracy (loss) over the validation set to the controller. The controller, then, optimizes its policy and updates its own parameters $w$ using the PQC feedback as well as characteristics of the distribution $\Pi_t(w)$, such as entropy. To do the update, controller uses the policy gradient method \cite{sutton2000policy}.

\begin{figure}
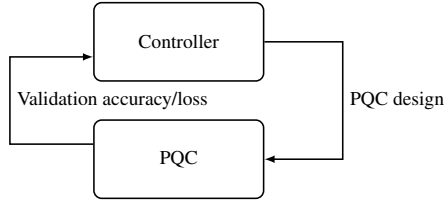

	\centering
	\includestandalone[width = 6cm]{Interaction}
	\caption{Controller-PQC interaction: the controller provide the PQC with a design and receives back the train accuracy (loss). Then, it updates its parameters by policy gradient RL method considering the PQC train report plus the suggested design entropy and policy distribution.}
	\label{Interaction}
\end{figure}

\begin{figure}
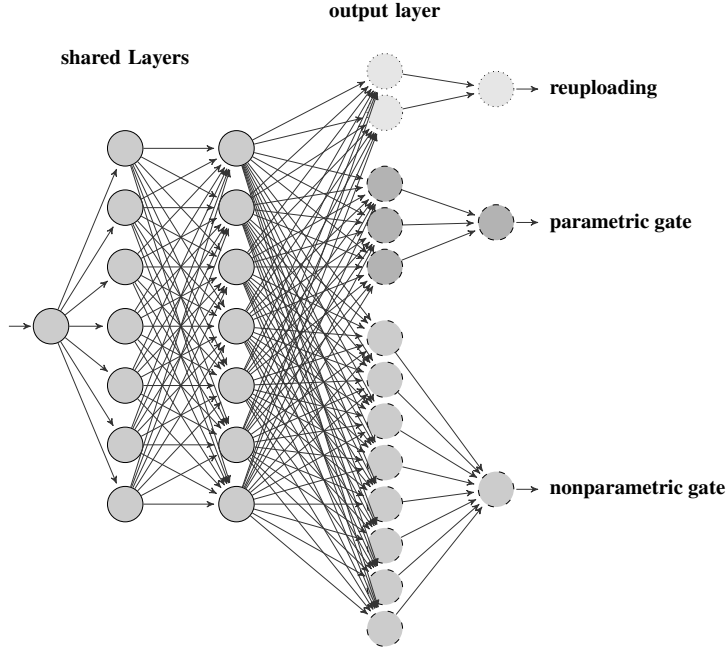

	\hspace*{-2cm}
	\centering
	\includestandalone[width = 10cm]{ControllerNetwork}
	\caption{Controller network structure: a deep neural network consisting of two shared layers for all the decisions plus a layer corresponding to each decision is considered. The output of the controller is a joint distribution of all the decisions from which we sample to find the design. }
	\label{ControllerNetwork}
\end{figure}

Figure \ref{Interaction} illustrates the environment interaction. Either controller suggests the entire design of the QPC or it only suggests the design of a layer which its pattern is repeated throughout the entire PQC. Figure \ref{ControllerNetwork} shows how controller presents the distribution over the decisions related to a single node (qubit). There are three decisions to make, uploading the data, parametric operation, nonparametric operation. In addition, PQC can send back the validation loss as the loss or validation accuracy as the reward to the controller. Therefore, we have two types of policies and two types of performance metrics resulting in the total of 4 different R-QCDS approaches. When we consider accuracy as the performance metric, as the loss for the controller we assume (1-accuracy). 

We use a feed-forward DNN in contrast to RNN networks used in \cite{zoph2016neural, pham2018efficient}. Therefore, there is no input to the controller. The controller has two shared layers (of the sizes of $48\times12$ in our study) and a specific layer for each of the reuploading, parametric and nonparametric gate decisions. The output layers provide us with the distributions over the corresponding decisions. Therefore, the number of nodes in the output layers matches the possible outcomes of corresponding decision. For instance, in the case of reuploading decision, the controller output is a Bernoulli distribution. Then, a sample is drawn and the desired action is suggested for the reuploading decision.

\begin{figure}
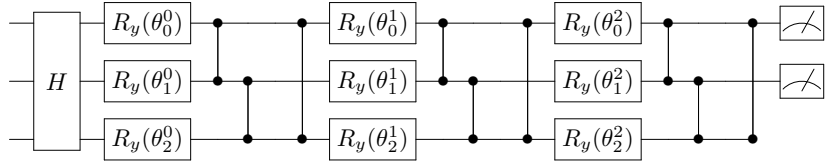

	\centering
	\includestandalone[width = 12cm]{smallGood}

	\caption{Hybrid quantum classical controller structure: a PQC corresponding to the decision of reuploading data.}
	\label{smallPQC}
\end{figure}

One can investigate the idea of having hybrid quantum classical controllers instead of entirely classical ones such as abovementioned DNNs. In this study, we consider separate small PQC as controllers to decide the structure of the original PQC. we develop PQCs with fixed structure to output the decisions for the original PQC design. In other words, we consider all parts of the R-QCDS framework to be hybrid algorithms. For instance, a simple 3 layer with 3 qubits PQC (of total 9 parameters) plus two single $\sigma_z$ measurements on the first two qubits is considered to decide whether we are reuploading the data. Figure \ref{smallPQC} illustrates this small PQC.

We consider smaller pool of candidates for the hybrid controller circuits due to computational expenses. We reduce the nonparametric choices to the set of \{CNOT, CZ, Tof, CSWAP\} but the rest are the same as the classical controller pool.

\subsection{Random Search}
Arguably the most simple yet powerful approach to search for a PQC design is a random search. We choose a large set of designs $D$ randomly. We further make sure that the selected designs are different enough from one another. That is, for every $d_1, d_2 \in D$, we check that their sequence of decisions match below a threshold considering Gestalt pattern matching \cite{ratcliff1988pattern}. We train all the designs for a very few epochs (in our numerical cases two) and keep some portion of the designs based on their loss (in our numerical cases the better half). Then, we train the remaining for few more epochs (five epochs in our studies) and repeat the process until we reach one of the following criteria: 1) We reduce the size of the remaining designs to a number where we can train them all to their best ability; 2) The number of training epochs experimentally reaches the best possible value; 3) We realize the rankings of the remaining designs with respect to their losses are not changing dramatically with increasing the training epochs after some initial steps.

At each layer for a single qubit we consider $2\times3\times8 =48$ possible outcomes having reuploading, parametric and nonparametric operations decisions. Consider a PQC with 6 layers of 4 qubits. The random search space ends up to have more than $10^{30}$ designs. We consider validation loss values for comparison between the random designs to rank them. One can alternatively use classification accuracy as the comparison metric.

\subsection{Black-Box Bayesian Optimization}
We focus on black-box Bayesian optimization (BO) in order to suggest new PQC designs. In this approach, we consider the PQC as a black-box function which we only have access to its objective function evaluations and we assign a prior Gaussian distribution to every PQC decisions' distribution. Gaussian processes inherently consider the decision variables to accept real values but here all the decision variables are discrete. We consider a hyper-rectangle $\{x \in \mathbb{R}^d | l_i \leq x_i \leq u_i\}$ as the feasible set of all decision variables with $l_i$ and $u_i$ be the lower and upper bounds for decision variable $x_i$. We divide the interval $\left[l_i,u_i\right]$ into unit length sub-intervals and assign one unique integer outcome to any of these sub-intervals similar to literature \cite{snoek2012practical,pirhooshyaran2020feature}. For instance, for $x_i$ representing a parametric gate option we set $l_i=-\frac{1}{2}$ and $u_i = \frac{5}{2}$ and then divide the interval into three sub-intervals as $\left[-\frac{1}{2},\frac{1}{2}\right]$, $\left[\frac{1}{2},\frac{3}{2}\right]$, $\left[\frac{3}{2},\frac{5}{2}\right]$ representing $R_x$, $R_y$ and $R_z$ respectively.

\begin{figure}
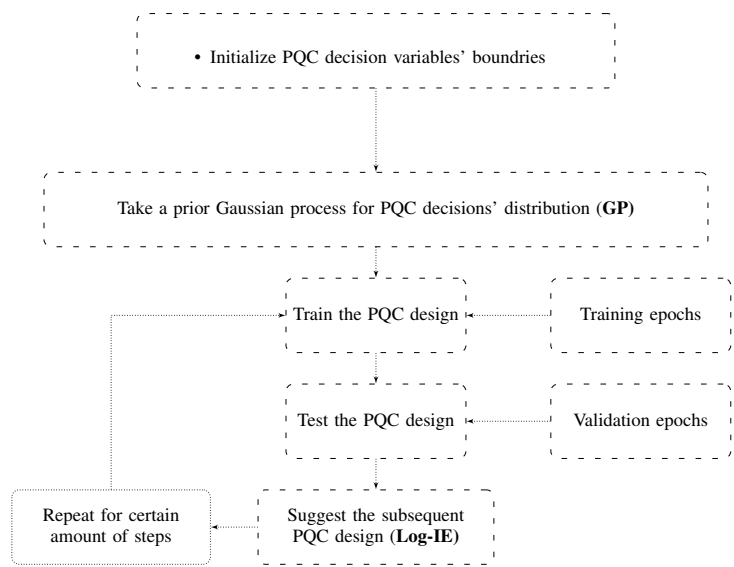

	\centering
	\includestandalone[width = 10cm]{GPEI}
	\caption{Gaussian process logarithmic expected improvement (GPLEI) Bayesian optimization method}
	\label{GPLEI}
\end{figure}

Figure \ref{GPLEI} illustrates the general BO process we use. We consider validation loss values as the comparison metric for BO to evaluate. We use logarithmic expected improvement (Log EI) as the acquisition function \cite{klein-bayesopt17}. That is, we aim to select the next design so that the reduction over the validation loss becomes as large as possible; however, any actual reduction is unknown until after the next examination. Therefore, we instead calculate the logarithmic expectation of the reduction and select the argmax of this expression.

\section{Results}
\label{results}
In this section, we provide numerical results and comparisons. During the experiments, we divide the datasets into three parts: train, validation and test unless otherwise mentioned. Models never see the test part. There are classical hyperparameters to be tuned in the learning process as well such as learning rate, mini-batch size. R-QCDS, also consists of other parameters such as controller learning rate, mini-batch size and entropy coefficient. To tune these parameters, we mainly use a very simple grid search. Throughout the numerical studies we consider the number of the circuit qubits to be equal to the features of the dataset. All the results are simulated via the Pennylane quantum machine learning library \cite{bergholm2018pennylane}.

\begin{figure}
	\centering
	\scalebox{0.35}{\input{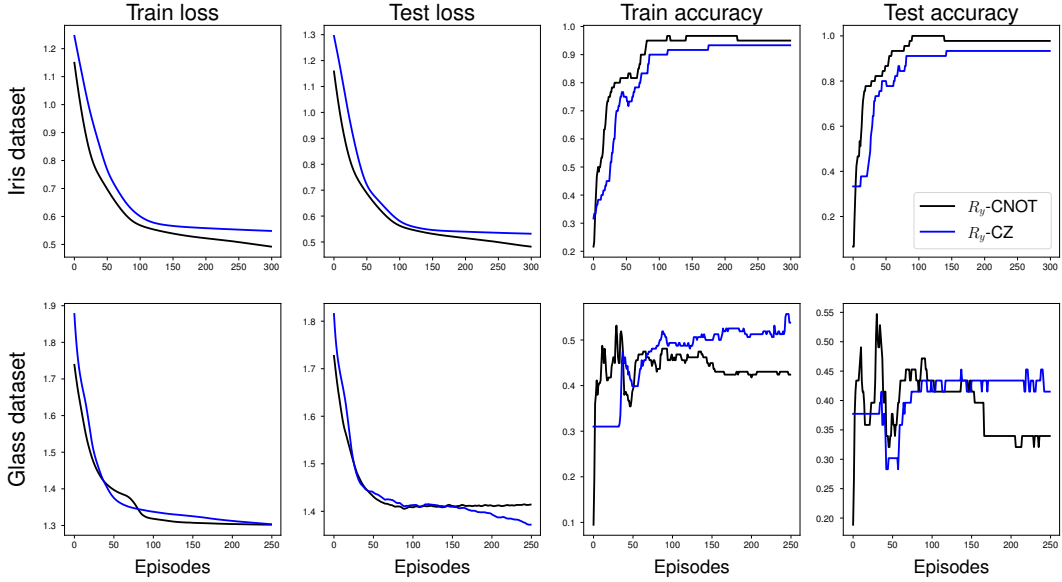}}
	\caption{Train loss, train accuracy, test loss and test accuracy comparisons for ($R_y$ rotation + CNOT) and ($R_y$ rotation + CZ) PQC designs on Iris and Glass datasets. ($R_y$ rotation + CNOT) design performs marginally better and hence would be considered as the benchmark for future comparisons. It reaches the test accuracy of $54.7\%$ for the Glass dataset utilizing $6\times 9$ parametric $R_y$ gates.}
	\label{TypicalComparison}
\end{figure}

In the following numerical studies we mainly use 6 quantum layers defined in the form of equation (\ref{eq:PQC}) stack upon each other. We use two datasets of Iris and Glass from \texttt{LIBSVM} repository. Both consist of multiple classes. Glass data points contain 9 features and 6 labels while Iris data points contain 4 features and 3 labels. First, we aim to show the validity of our modeling by illustrating the performances of two already established designs in the literature. Hence, we consider a pair of a single rotation $R_y$ plus CNOT and a single rotation $R_y$ plus CZ for each qubit \cite{mcclean2018barren,cerezo2020cost,grant2019initialization}. Figure \ref{TypicalComparison} shows the performances of the two designs. The train and test losses are divided by the size of their datasets (and for the rest of the studies as well). For the Iris dataset the ($R_y$ rotation + CNOT) design reaches 100\% test accuracy at around 100 epochs into the training process. We can also see that ($R_y$ rotation + CNOT) design works marginally better than ($R_y$ rotation + CZ).
 
\begin{figure}
	\hspace*{-1cm}
	\includegraphics[width = 18cm]{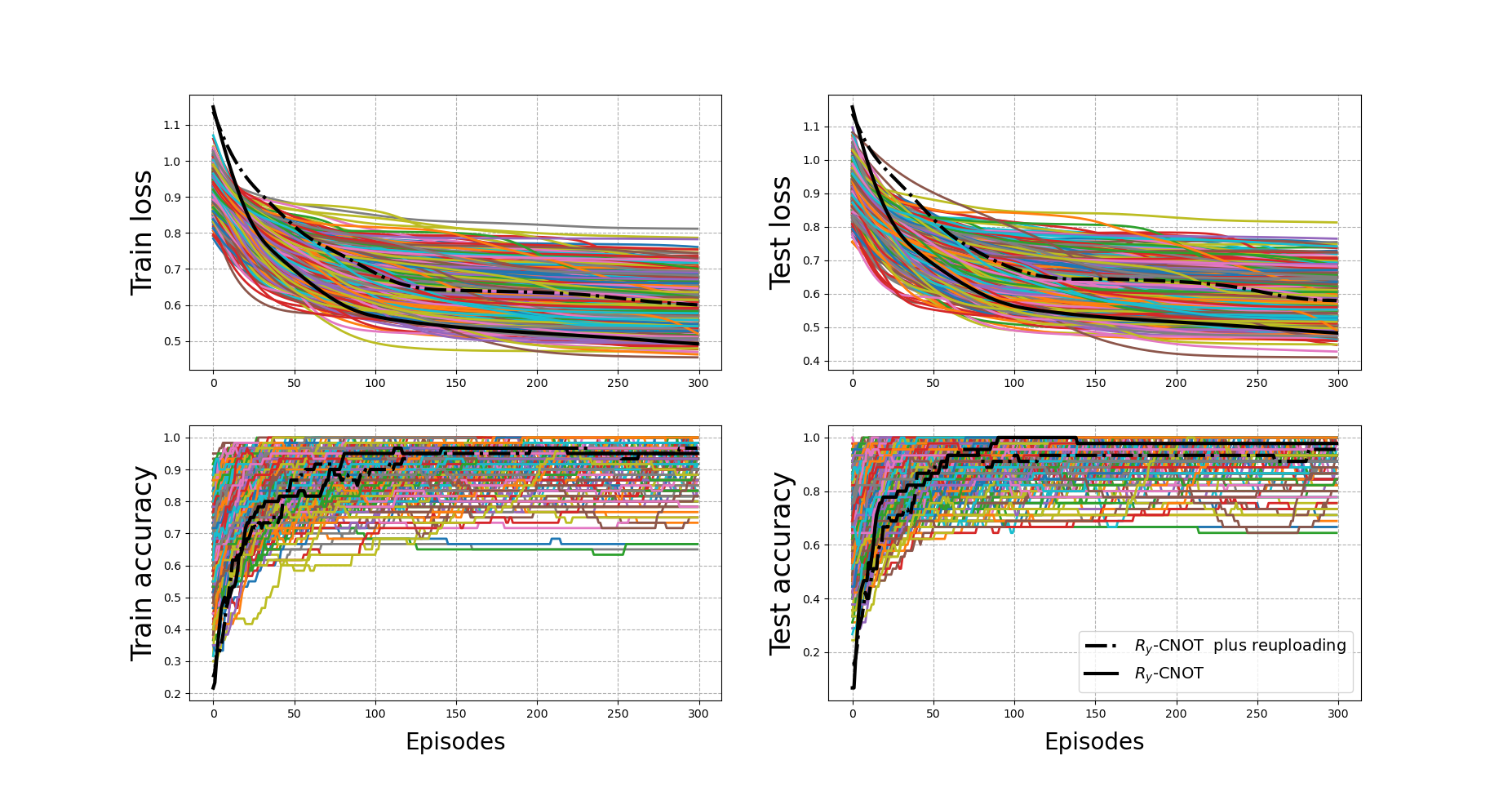}
	\caption{Train loss, train accuracy, test loss and test accuracy comparisons between the best 1000 designs found by random search and $R_y$ rotation + CNOT design. One can certainly see some designs outperform the benchmark. In terms of test loss comparison, 2.5\% of the set of the best 1000 random designs perform considerably better than benchmark. For the best 1000 discovered designs one can see the paper's \href{https://github.com/mamadpierre/QCDS}{repository}}
	\label{RandomSearchComparison}
\end{figure}

We, therefore, consider ($R_y$ rotation + CNOT) as the benchmark for comparison with designs extracted from the three proposed QCDS methods. First, we consider the random search method. We create 30 thousand randomly chosen designs. We do not allow designs to posses sequences matched more than 75\%. We train them only for two epochs and rank them based on validation loss and throw away the worst half. Subsequently, we repeat the process considering the remaining designs trained for 5 epochs and then for 10 epochs. After reaching this point, we realize that the ranking of the best 1000 random designs doesn't change dramatically. Hence, we stop the process and train all the best 1000 designs for 300 epochs. Figure \ref{RandomSearchComparison} illustrates the comparison between the benchmark design and the 1000 best designs found by the random search on the Iris dataset. We consider 40\% training, 30\% validation and 30\% test.  

As can be seen in Figure \ref{RandomSearchComparison}, many designs fail to produce competitive results but some of the discovered architectures perform considerably better than the ($R_y$ rotation + CNOT) design in terms of train and test losses. Although the benchmark reaches the test accuracy of 100\% as discussed before, many randomly discovered designs reach the same accuracy much quicker and also these designs are mainly robust enough to stay at 100\% accuracy unlike the benchmark for which the test accuracy drops slightly. Considering that we conducted this study on a small- to moderate-sized dataset where the benchmark by itself reaches 100\% accuracy as well as we only created 30k designs which is very tiny fraction of the designs for the considered PQC, this results reassure that there are many powerful and trainable designs in the search space.

We also investigate the performance of the benchmark when at each layer for all the qubits, we reupload the initial data. This has been shown as $R_y$-CNOT plus reuploading in Figure \ref{RandomSearchComparison}. One can see that this design performs inferior compared with the benchmark which in a sense emphasizes the fact that stacking up more gates without careful thought or any optimizing mechanism can easily result in less trainable quantum designs. One should only consider the introduced reuploading strategy as an option which based on an optimization strategy can be switched off or on for different qubits in different layers.

\begin{figure}
	\hspace*{-1.5cm}
	\scalebox{0.40}{\input{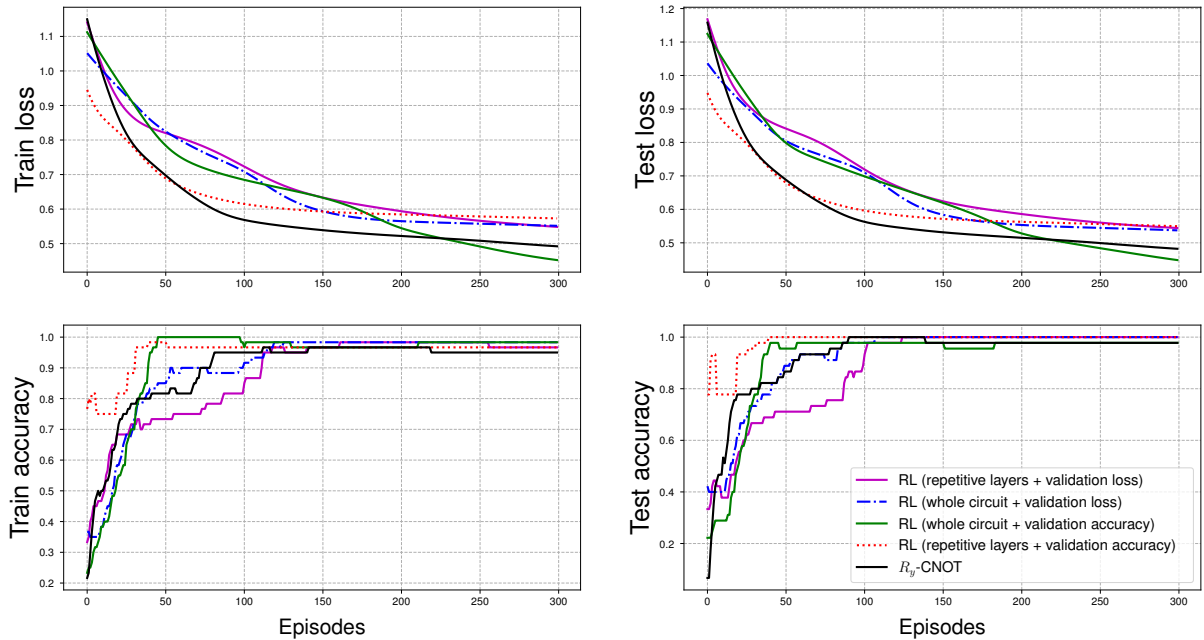}}
	\caption{Train loss, train accuracy, test loss and test accuracy comparisons between R-QCDS strategies and typical $R_y$ rotation + CNOT design. We consider four R-QCDS approaches. We use validation loss or validation accuracy as the performance metric. Moreover, we allow the R-QCDS methods to suggest only a quantum layer or the whole circuit. Controller learning rate is $0.1$, $0.2$, $0.02$ and $0.02$ for the different R-QCDS approaches in the order written in the plot legend.}
	\label{RQCDSComparison}
\end{figure}

Now, we focus on the R-QCDS approaches discussed in the Section \ref{RQCDS}. For each feedback loop we train the suggested PQC from scratch for 50 epochs. After less than 50 feedback loop (depending on how large the controller learning rate is), controller's suggestions converge to a unique design. Figure \ref{RQCDSComparison} illustrates the comparison between 4 separate RL schemes and the bench mark ($R_y$ rotation + CNOT). The Policy gradient methods are stable in terms of train and test accuracies but they are marginally inferior in terms of test loss in comparison with the benchmark. 

\begin{figure}
	\centering
	\scalebox{0.30}{\input{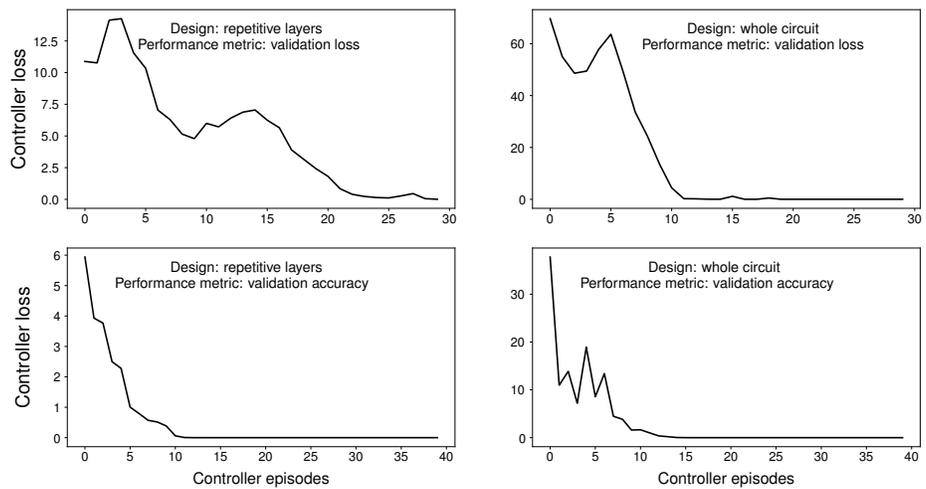}}
	\caption{Controller's learning curve for 4 different R-QCDS strategies on the validation dataset.}
	\label{ControllerLoss}
\end{figure}

Figure \ref{ControllerLoss} illustrates the controller's learning curve on the validation dataset. For each point shown in the plots, a design is sent to the PQC, PQC is trained for 50 epochs and then the validation loss or accuracy on the validation dataset is fed back to the controller, then the policy plus entropy loss of the controller based on policy gradient RL method is calculated and plotted.     

\begin{figure}
		\centering
	\scalebox{0.35}{\input{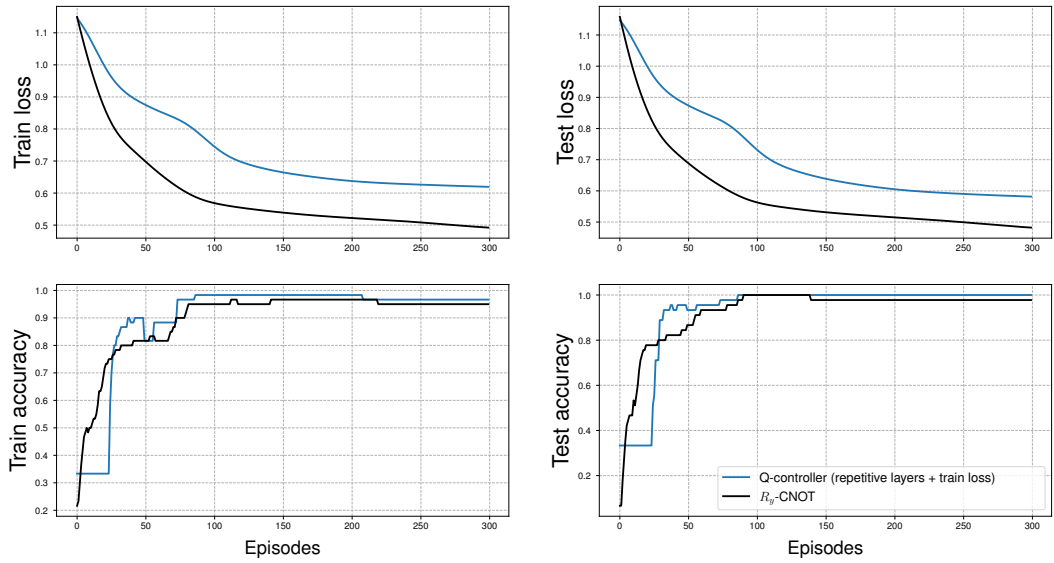}}
	\caption{Train loss, train accuracy, test loss and test accuracy comparisons for ($R_y$ rotation + CNOT) and Q-controller suggested circuits on IRIS dataset.}
	\label{fig:Q-controller}
\end{figure}

We further consider the case of having hybrid classical quantum controller for R-QCDS. We only consider the case of controller suggesting repetitive layers. We consider 3-, 3- and 4-qubit PQCs of 3 layers with 2, 2 and 4 measurements at the end for reuploading, rotation and nonparametric decisions respectively and for each of the original PQC qubits. We illustrated one such small PQC controller in Figure \ref{smallPQC} in the previous section. 12 small PQCs are considered to design a single layer of the original PQC. Figure \ref{fig:Q-controller} illustrates the train and test comparisons between the suggested design by hybrid controller R-QCDS (based on loss metric) and the benchmark on the IRIS dataset.

\begin{figure}
		\centering
	\scalebox{0.35}{\input{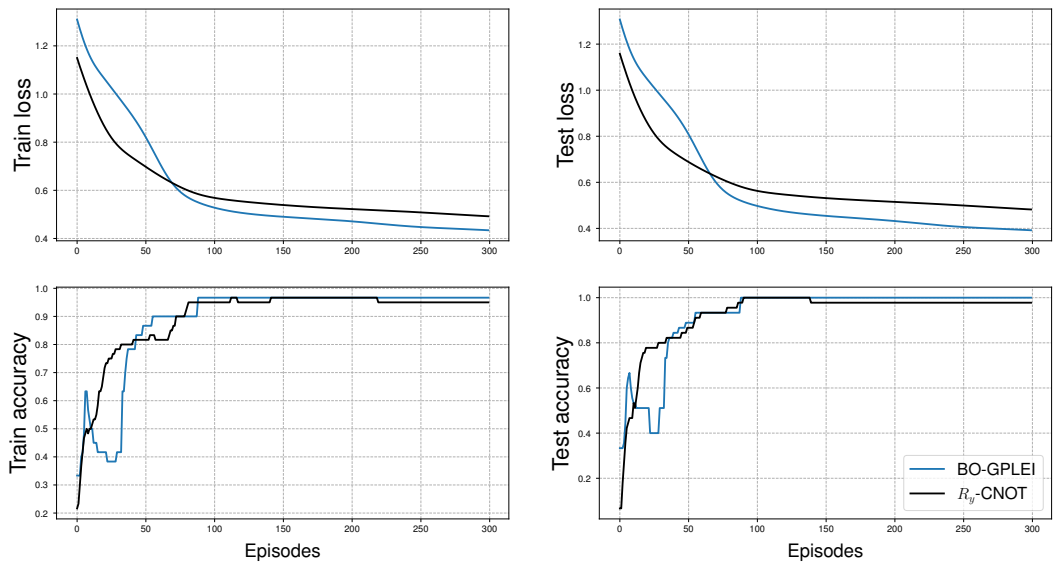}}
	\caption{Train loss, train accuracy, test loss and test accuracy comparisons between GPLEI Bayesian optimization and $R_y$ rotation + CNOT design.}
	\label{RoBOComparison}
\end{figure}

At last, we discuss the BO results. As we mentioned earlier, more accurate the segregate function results in better performance of BO. Hence, we train the PQCs for 100 epochs at each time step to then provide the validation loss. After around 70 function evaluations, BO suggests a design which performs superior to the ($R_y$ rotation + CNOT) design. So, we stop the process. 
Figure \ref{RoBOComparison} illustrates the comparison.  

\subsection{Generalization}
The previous findings on Iris dataset confirm the importance of the QCDS. The superior designs in terms of trainability may inherently be more efficient than the benchmark and therefore extremely desirable for other applications and/or datasets. So, we go one step further and aim to compare the most efficient designs found in the previous section with the benchmark on Glass dataset. We put an emphasis on the fact that the selected designs are totally agnostic towards Glass dataset. In addition, the Glass dataset consists of 9 features for each data point and we assign a PQC of side 9 qubits to it. However, the architectures introduced previously consist of four qubits. So, we simply repeat the pattern to fully cover the Glass PQC. For instance, considering $0,\dots, 8$ qubits used in the PQC for the Glass dataset, 8th and 4th qubits have similar operations as 0th qubit has with new tunable parameters.

\begin{figure}
	\hspace*{-2.5cm}
	\scalebox{0.45}{\input{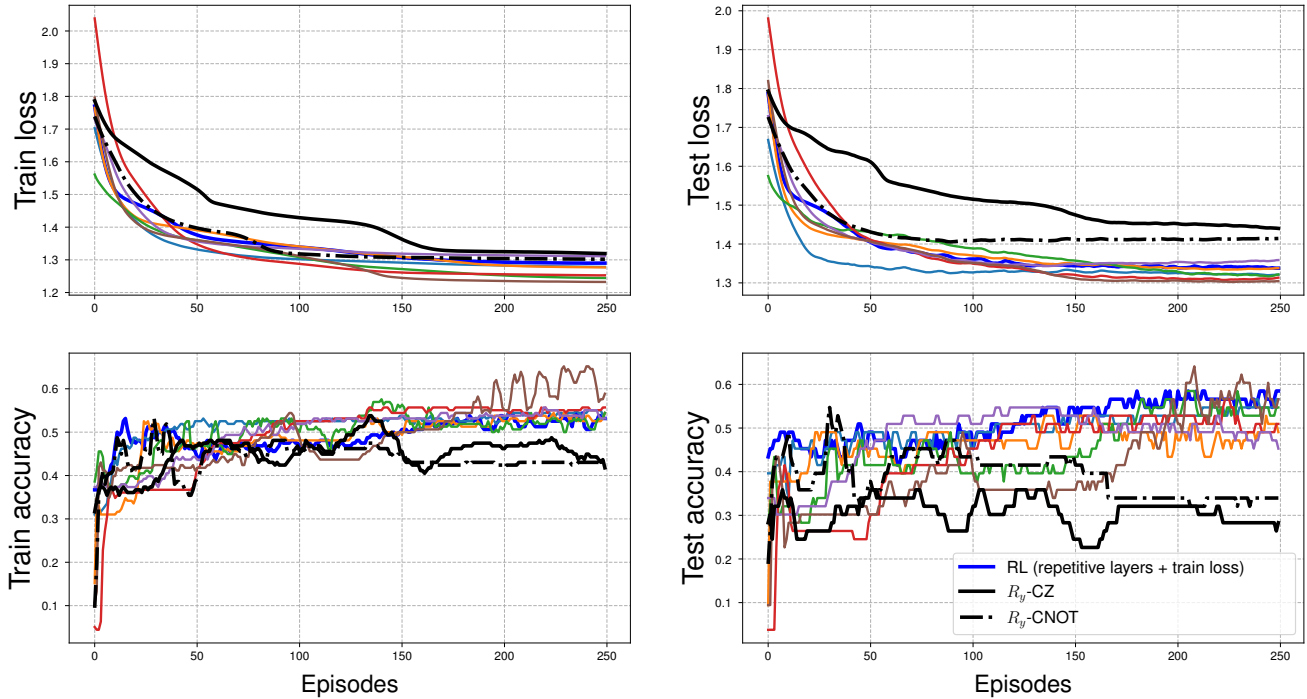}}
	\caption{Train loss, train accuracy, test loss and test accuracy comparisons between selected designs and the benchmark design on Glass dataset. The second best test accuracy result is achieved by the design discovered by R-QCDS (design: repetitive layers plus performance metric: validation loss). The best accuracy reported is $64.1\%$ from a randomly discovered design which presents $9.4\%$ increase over the benchmark design test accuracy $54.7$. In total, 6 designs report test accuracy increases. For their designs one can see the paper's \href{https://github.com/mamadpierre/QCDS}{repository}.}
	\label{IrisToGlass}
\end{figure}

\begin{sidewaysfigure}
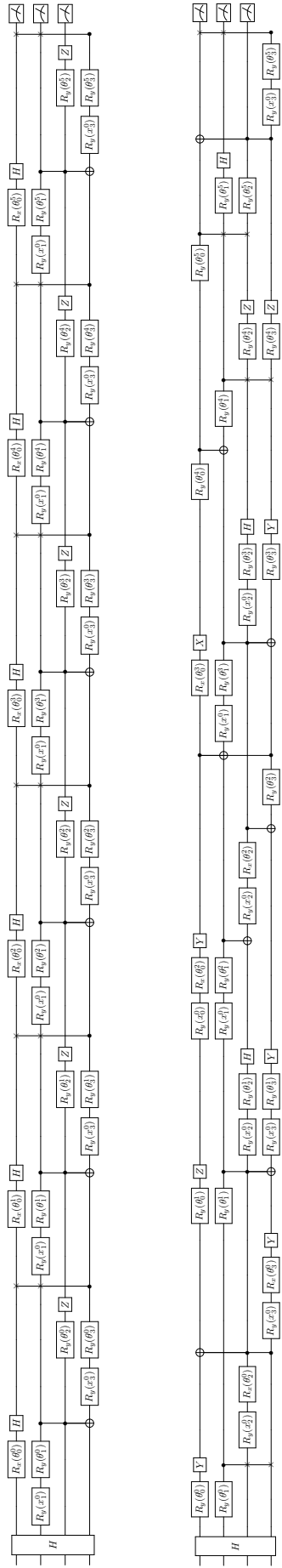

	\hspace*{-0.5cm}
	\includestandalone[width = 25cm]{GoodDesign0}

	\\\\\\
	
	\hspace*{-0.5cm}
	\includestandalone[width = 25cm]{GoodDesign20}

	\caption{The best two designs reported in Figure \ref{IrisToGlass}. The top section design is suggested by R-QCDS (design: repetitive layers plus performance metric: validation loss) while the bottom one is one of the randomly chosen designs which performed better than the benchmark on the Iris dataset. In terms of test accuracy they outperform the benchmark by reporting $58.4\%$ and $64.1\%$ accuracies, respectively. The designs are originally optimized to be used on Iris dataset. So, readers can see PQCs consists of 4 qubits with three single qubit $\sigma_z$ measurements at the end.}
	\label{IrisToGlassExamples}
\end{sidewaysfigure}

We consider randomly chosen designs that performed considerably better than the benchmark, plus designs suggested by four different R-QCDS strategies plus BO. In this experiment, we split the data into 75\% train and 25\% test. No validation dataset is needed because the designs are already fixed. Methods never see the test dataset during the training. Although selected architectures were solely designed based on their performance on Iris dataset, the results shown in \ref{IrisToGlass} strongly suggest they perform meaningfully better than the benchmark on another unseen dataset. The second best test accuracy result is achieved by a design discovered by R-QCDS (repetitive layers + train loss) and the best test accuracy result is achieved by one of the randomly discovered designs. The best test accuracy reported is $64.1\%$ which means $9.4\%$ increase over the benchmark design test accuracy $54.7$. Figure \ref{IrisToGlassExamples} illustrates these two designs.
  
\section{Conclusion} 
\label{Conclusion}
This paper represents one of the first research which focuses on quantum circuit design search framework considering trainability of parameterized quantum circuits. Three optimization strategies are provided to find more efficient designs: random search plus survival of the fittest; reinforcement learning and Bayesian optimization. We suggest nontrivial automated designs which are hard to be hand-designed. We investigate their performances on unseen dataset and compare the results with benchmark parameterized quantum circuit designs. One can see the strong impact of optimization strategies in designing of parameterized quantum circuits.

\bibliography{QCDS}

\bibliographystyle{unsrtnat}

\end{document}

%% file: Interaction.tex
	\tikzstyle{block} = [rectangle, draw, 
	text width=8em, text centered, rounded corners, minimum height=4em]
	
	\tikzstyle{line} = [draw, -latex]
	
	\begin{tikzpicture}[node distance = 6em, auto, thick]
		\node [block] (Agent) {Controller};
		\node [block, below of=Agent] (Environment) {PQC};
		
		\path [line] (Agent.0) --++ (4em,0em) |- node [near start]{PQC design} (Environment.0);
		\path [line] (Environment.170) --++ (-4.25em,0em) |- node [near start, right] {{Validation accuracy/loss}} (Agent.190);
	\end{tikzpicture}

%% file: ControllerNetwork.tex
\pagestyle{empty}

\def\layersep{2.5cm}

\begin{tikzpicture}[shorten >=2pt,->,draw=black!80, node distance=\layersep]
    \tikzstyle{every pin edge}=[<-,shorten <=1pt]
    \tikzstyle{neuron}=[circle,fill=black!25,minimum size=17pt,inner sep=0pt]
    \tikzstyle{neuron2}=[circle,fill=black!25,minimum size=17pt,inner sep=0pt]
    \tikzstyle{input neuron}=[neuron, draw=black,fill=gray!40];
    \tikzstyle{input neuron2}=[neuron2, draw=black,fill=gray!40];
    \tikzstyle{output neuron}=[neuron, dotted,draw=black,fill=gray!20];
    \tikzstyle{output neuron2}=[neuron, dashed,draw=black,fill=gray!60];
    \tikzstyle{output neuron3}=[neuron, loosely dashed,draw=black,fill=gray!40];
    \tikzstyle{hidden neuron}=[neuron, dotted,draw=black,fill=gray!20];
    \tikzstyle{hidden neuron2}=[neuron, dashed,draw=black,fill=gray!60];
    \tikzstyle{hidden neuron3}=[neuron, loosely dashed,draw=black,fill=gray!40];
    \tikzstyle{annot} = [text width=8em, text centered]
    \tikzstyle{annot2} = [text width=8em, ,text centered]
    
        \node[input neuron, pin=left:] (I-1) at (0,-4.5) {};

    \foreach \name / \y in {1,...,7}
		\node[input neuron2] (II-\name) at (0.5 * \layersep,-\y-0.5) {};

    \foreach \name / \y in {1,...,7}
	\node[input neuron2] (III-\name) at (1.25 *\layersep,-\y-0.5) {};

    \foreach \name / \y in {1,2}
        \path[yshift=0.5cm]
            node[hidden neuron] (H-\name) at (2.25*\layersep,-0.7*\y ) {};
    \foreach \name / \y in {3,...,5}
		\path[yshift=0.5cm]
			node[hidden neuron2] (H-\name) at (2.25*\layersep,-0.7*\y -0.5 ) {};

    \foreach \name / \y in {6,...,13}
		\path[yshift=0.5cm]
			node[hidden neuron3] (H-\name) at (2.25*\layersep,-0.7*\y -1 ) {};

    \node[output neuron,pin={[pin edge={->}]right:\bf reuploading}] (O) at (3*\layersep, -0.5 ) {};
    \node[output neuron2,pin={[pin edge={->}]right:\bf parametric gate}] at (3*\layersep, -2.75) (OO) {};
    
    \node[output neuron3,pin={[pin edge={->}]right:\bf nonparametric gate}] (OOO) at (3*\layersep, -7.25) {};

    \foreach \source in {1}
		\foreach \dest in {1,...,7}
			\path (I-\source) edge (II-\dest);    

    \foreach \source in {1,...,7}
		\foreach \dest in {1,...,7}
			\path (II-\source) edge (III-\dest);

    \foreach \source in {1,...,7}
        \foreach \dest in {1,...,13}
            \path (III-\source) edge (H-\dest);

    \foreach \source in {1,2}
        \path (H-\source) edge (O);
        
    \foreach \source in {3,...,5}
		\path (H-\source) edge (OO);        
        
	\foreach \source in {6,...,13}
		\path (H-\source) edge (OOO);

    \node[annot,above of=II-1, node distance=1.5cm] (hl2) {\bf shared Layers}; 
    \node[annot,above of=H-1, node distance=1cm] (hl2) {\bf output layer};                  
\end{tikzpicture}

%% file: smallGood.tex
\hspace{1em}	
\begin{tabular}{c}
	\vspace{-.6em}\\
\Qcircuit @C=1em @R=.7em {
	& \multigate{2}{H}&   \gate{R_y(\theta_0^0)} &\ctrl{1}  &  \qw &  \qw&  \ctrl{2}
	&  \gate{R_y(\theta_0^1)} &\ctrl{1}  &  \qw &  \qw&  \ctrl{2}
	&  \gate{R_y(\theta_0^2)} &\ctrl{1}  &  \qw &  \qw&  \ctrl{2} &\meter\\
	&\ghost{H}		&  \gate{R_y(\theta_1^0)} &\ctrl{-1}  &  \ctrl{1} &  \qw&  \qw
	&\gate{R_y(\theta_1^1)} &\ctrl{-1}  &  \ctrl{1} &  \qw&  \qw
	&  \gate{R_y(\theta_1^2)} &\ctrl{-1}  &  \ctrl{1} &  \qw&  \qw  &\meter \\
	&\ghost{H}    &   \gate{R_y(\theta_2^0)} & \qw &  \ctrl{-1} &  \qw&  \ctrl{-2}
	&  \gate{R_y(\theta_2^1)} &\qw  &  \ctrl{-1} &  \qw&  \ctrl{-2}
	&  \gate{R_y(\theta_2^2)} &\qw  &  \ctrl{-1} &  \qw&  \ctrl{-2} 
}

	\vspace{1.2em}\hspace{1.2em}
	\\
\end{tabular}

%% file: GPEI.tex
	
\tikzstyle{decision} = [diamond,loosely dashed,draw=black,fill=gray!40,draw, fill=white!20, text width=6em,  text centered,  node distance=3cm, inner sep=0pt]
\tikzstyle{block0}   = [rectangle, loosely dashed,draw=black,fill=gray!40,draw, fill=white!20, text width=25em, text centered,  node distance=3cm, rounded corners, minimum height=4em]
\tikzstyle{block00}  = [rectangle, loosely dashed,draw=black,fill=gray!40,draw, fill=white!20, text width=35em, text centered,  node distance=3cm, rounded corners, minimum height=4em]
\tikzstyle{block}    = [rectangle, loosely dashed,draw=black,fill=gray!40,draw, fill=white!20, text width=9em,  text centered,                     rounded corners, minimum height=4em]
\tikzstyle{block1}   = [rectangle, loosely dashed,draw=black,fill=gray!40,draw, fill=white!20, text width=12em, text centered,                     rounded corners, minimum height=4em]
\tikzstyle{line}     = [densely dotted, draw, -latex']
\tikzstyle{invisibleblock}    = [rectangle, densely dotted,draw=black,fill=gray!40,fill=white!20, text width=10em,  text centered,  node distance=3cm,    rounded corners, minimum height=4em]

  \centering

  \begin{tikzpicture}[node distance = 2cm, auto]
    \node [block0, align=center] (init) { \begin{itemize} \item Initialize PQC decision variables' boundries \end{itemize}};
    \node [block00, align=center,below of = init] (GP) {Take a prior Gaussian process for PQC decisions' distribution (\bf GP)};
    \node [block, below of = GP, node distance = 2cm] (yes block1) {Train the PQC design};
    \node [block, below of = yes block1, node distance = 2cm] (yes block2) {Test the PQC design};
    \node [block1, below of = yes block2, node distance = 2cm] (yes block4) {Suggest the subsequent PQC design (\bf Log-IE)};
    \node [block, right of = yes block1, node distance = 5cm] (Tr) {Training epochs};
	\node [block, right of = yes block2, node distance = 5cm] (Val) {Validation epochs};
	\node [invisibleblock, left of = yes block4, node distance =5cm](invis){Repeat for certain amount of steps};
    \path [line] (init) -- (GP);
    \path [line] (GP) -- node {}(yes block1);
    \path [line] (yes block1) -- node {}(yes block2);
    \path [line] (yes block2) -- node {}(yes block4);    
    \path [line] (yes block4) -- node {}(invis);
     \path [line] (invis) |- node {}(yes block1);           
    \path [line] (Tr) -- node {}(yes block1);
    \path [line] (Val) -- node {}(yes block2);

  \end{tikzpicture}

%% file: GoodDesign0.tex
\hspace{1em}	
\begin{tabular}{c}
	\vspace{-.6em}\\
\Qcircuit @C=1em @R=.7em {
	& \multigate{3}{H}&  \qw &  \gate{R_x(\theta_0^0)} &\gate{H}  &  \qw &  \qw&  \qw&\qswap
	&  \qw &  \gate{R_x(\theta_0^1)} &\gate{H}  &  \qw &  \qw&  \qw&\qswap
	&  \qw &  \gate{R_x(\theta_0^2)} &\gate{H}  &  \qw &  \qw&  \qw&\qswap
	&  \qw &  \gate{R_x(\theta_0^3)} &\gate{H}  &  \qw &  \qw&  \qw&\qswap
	&  \qw &  \gate{R_x(\theta_0^4)} &\gate{H}  &  \qw &  \qw&  \qw&\qswap
	&  \qw &  \gate{R_x(\theta_0^5)} &\gate{H}  &  \qw &  \qw&  \qw&\qswap &\meter\\
	&\ghost{H}		  &  \gate{R_y(x_1^0)} &  \gate{R_y(\theta_1^0)} &\ctrl{2} &   \qw&  \qw&   \qw&\qswap
	&  \gate{R_y(x_1^0)} &  \gate{R_y(\theta_1^1)} &\ctrl{2} &   \qw&  \qw&   \qw&\qswap
	&  \gate{R_y(x_1^0)} &  \gate{R_y(\theta_1^2)} &\ctrl{2} &   \qw&  \qw&   \qw&\qswap
	&  \gate{R_y(x_1^0)} &  \gate{R_y(\theta_1^3)} &\ctrl{2} &   \qw&  \qw&   \qw&\qswap
	&  \gate{R_y(x_1^0)} &  \gate{R_y(\theta_1^4)} &\ctrl{2} &   \qw&  \qw&   \qw&\qswap
	&  \gate{R_y(x_1^0)} &  \gate{R_y(\theta_1^5)} &\ctrl{2} &   \qw&  \qw&   \qw&\qswap  &\meter \\
	&\ghost{H}        &\qw  &\qw   &\ctrl{1} &  \qw &  \gate{R_y(\theta_2^0)} &\gate{Z}  &  \qw
	&\qw  &\qw   &\ctrl{1} &  \qw &  \gate{R_y(\theta_2^1)} &\gate{Z}  &  \qw
	&\qw  &\qw   &\ctrl{1} &  \qw &  \gate{R_y(\theta_2^2)} &\gate{Z}  &  \qw
	&\qw  &\qw   &\ctrl{1} &  \qw &  \gate{R_y(\theta_2^3)} &\gate{Z}  &  \qw
	&\qw  &\qw   &\ctrl{1} &  \qw &  \gate{R_y(\theta_2^4)} &\gate{Z}  &  \qw
	&\qw  &\qw   &\ctrl{1} &  \qw &  \gate{R_y(\theta_2^5)} &\gate{Z}  &  \qw &\meter \\
	&\ghost{H}        &\qw  &\qw   &\targ &  \gate{R_y(x_3^0)} &  \gate{R_y(\theta_3^0)} & \qw &\ctrl{-3}
	&\qw  &\qw   &\targ &  \gate{R_y(x_3^0)} &  \gate{R_y(\theta_3^1)} & \qw &\ctrl{-3}
	&\qw  &\qw   &\targ &  \gate{R_y(x_3^0)} &  \gate{R_y(\theta_3^2)} & \qw &\ctrl{-3}
	&\qw  &\qw   &\targ &  \gate{R_y(x_3^0)} &  \gate{R_y(\theta_3^3)} & \qw &\ctrl{-3}
	&\qw  &\qw   &\targ &  \gate{R_y(x_3^0)} &  \gate{R_y(\theta_3^4)} & \qw &\ctrl{-3}
	&\qw  &\qw   &\targ &  \gate{R_y(x_3^0)} &  \gate{R_y(\theta_3^5)} & \qw &\ctrl{-3} 
}

	\vspace{1.2em}\hspace{1.2em}
	\\
\end{tabular}

%% file: GoodDesign20.tex
\hspace{1em}	
\begin{tabular}{c}
	\vspace{-.6em}\\
\Qcircuit @C=1em @R=.7em {
	&\multigate{3}{H}&\qw&\gate{R_y(\theta_0^0)}&\gate{Y}&\qw&\qw&\targ&\qw&\qw&\qw&\gate{R_y(\theta_0^1)}&\gate{Z}&\qw&\qw&\qw&\gate{R_y(x_0^0)}&\gate{R_x(\theta_0^2)}&\gate{Y}&\qw&\qw&\qw&\qw&\qw&\ctrl{1}&\qw&\gate{R_x(\theta_0^3)}&\gate{X}&\qw&\qw&\qw&\qw&\gate{R_y(\theta_0^4)}&\ctrl{1}&\qw&\qw&\qw&\qw&\qw&\qw&\qw&\gate{R_y(\theta_0^5)}&\ctrl{2}&\qw&\qw&\qw&\targ &\qw&\qw&\qswap &\meter\\
	&\ghost{H}&\qw&\gate{R_y(\theta_1^0)}&\ctrl{2}&\qw&\qw&\qw&\qw&\qw&\qw&\gate{R_y(\theta_1^1)}&\ctrl{2}&\qw&\qw&\qw&\gate{R_y(x_1^0)}&\gate{R_y(\theta_1^2)}&\ctrl{1}&\qw&\qw&\qw&\qw&\qw&\targ&\gate{R_y(x_1^0)}&\gate{R_y(\theta_1^3)}&\ctrl{2}&\qw&\qw&\qw&\qw&\qw&\targ&\qw&\gate{R_y(\theta_1^4)}&\ctrl{2}&\qw&\qw&\qw&\qw&\qw&\qswap&\qw&\gate{R_y(\theta_1^5)}&\gate{H}&\qw&\qw&\qw&\qswap &\meter\\
	&\ghost{H}&\qw&\qw&\qswap&\gate{R_y(x_2^0)}&\gate{R_x(\theta_2^0)}&\ctrl{-2}&\qw&\qw&\qw&\qw&\ctrl{1}&\gate{R_y(x_2^0)}&\gate{R_y(\theta_2^1)}&\gate{H}&\qw&\qw&\targ&\gate{R_y(x_2^0)}&\gate{R_x(\theta_2^2)}&\ctrl{1}&\qw&\qw&\qw&\qw&\qw&\ctrl{1}&\gate{R_y(x_2^0)}&\gate{R_y(\theta_2^3)}&\gate{H}&\qw&\qw&\qw&\qw&\qw&\qswap&\qw&\gate{R_y(\theta_2^4)}&\gate{Z}&\qw&\qw&\qswap&\qw&\gate{R_y(\theta_2^5)}&\qw&\ctrl{-2}&\qw&\qw&\qw &\meter\\
	&\ghost{H}&\qw&\qw&\qswap&\qw&\qw&\ctrl{-3}&\gate{R_y(x_3^0)}&\gate{R_x(\theta_3^0)}&\gate{Y}&\qw&\targ&\gate{R_y(x_3^0)}&\gate{R_y(\theta_3^1)}&\gate{Y}&\qw&\qw&\qw&\qw&\qw&\targ&\qw&\gate{R_y(\theta_3^2)}&\ctrl{-3}&\qw&\qw&\targ&\qw&\gate{R_y(\theta_3^3)}&\gate{Y}&\qw&\qw&\qw&\qw&\qw&\qswap  &\qw&\gate{R_y(\theta_3^4)}&\gate{Z}&\qw&\qw&\qw&\qw&\qw&\qw&\ctrl{-3}&\gate{R_y(x_3^0)}&\gate{R_y(\theta_3^5)}&\ctrl{-3}
}
	\vspace{1.2em}\hspace{1.2em}
	\\
\end{tabular}